# Goldstone Apple Valley Radio Telescope Observations of 2012 Solar Eclipse: A Multi-wavelength study of cm-λ Gyroresonance Emission from Active Regions

T. Velusamy[1], T. B. H Kuiper[1], S. M. Levin[1]
R. Dorcey[2], N. Kreuser-Jenkins[2], J. Leflang[2]

[1] Jet Propulsion Laboratory, California Institute of Technology, Pasadena, CA 91109, USA
[2] Lewis Center for Educational Research, 17500 Mana Road, Apple Valley, CA 92307

Corresponding Author: Thangasamy Velusamy
Email: thangasamy.velusamy@jpl.nasa.gov

**Abstract**

Goldstone Apple Valley Radio Telescope (GAVRT) is a science education partnership among NASA, the Jet Propulsion Laboratory (JPL), and the Lewis Center for Educational Research (LCER), offering unique opportunities for K -12 students and their teachers. The GAVRT program operates a 34-m radio telescope with a wide-band, low noise receiver, which is tunable in four independent dual-polarization bands from 3 to 14 GHz.  The annular eclipse of the Sun on 2012 May 20 was observed by GAVRT as part of education outreach.  In this paper we present the results of this eclipse data and discuss the multi-wavelength strip scan brightness distribution across three active regions.  We derive the source brightness temperatures and angular sizes as a function of frequency and interpret the results in terms of the gyroresonance mechanism.  We show examples of the increasing brightness and widening of source size (isogauss surface) with wavelength as evidence for gyroresonance emission layers of broader (diverging ) isogauss surfaces of the magnetic field geometry in the corona above solar surface. We present an example how the derived frequency – brightness temperature relationship is translated to a magnetic field – brightness temperature relationship under the frame-work of gyroresonance emission.  Our results demonstrate the usefulness of GAVRT bands as excellent probes to study the layers of the corona above the active regions (sun spots), in particular the prevalence of the gyroresonance mechanism.  Our results provide a frame-work for multiwavelength cm-λ eclipse observations and illustrate how the GAVRT program and K- 12 student/teacher participation can produce science data useful to the scientific community and science missions.

## 1. Introduction

Maps of the Sun's radio emission at centimeter-wavelengths (cm-λ) are dominated by the gyro-resonant emission above solar active regions produced by hot electrons gyrating in the coronal magnetic field (Kundu, 1965; Zhelesnyakov, 1970).  When the gyro-resonance mechanism for a given harmonic produces high enough opacity it measures the "local" temperature and magnetic field at the corresponding coronal height.  It is widely accepted that multi-wavelength, cm-λ data can provide diagnostics to a wide range of coronal and heights (c.f. Kundu. 1982).  In view of the recent developments in the measurement of photospheric magnetic fields and techniques for their extrapolation into the solar corona, the cm-λ radio measurements have become essential to enhance the extrapolations since only through comparison with actual data can those techniques be validated (c.f. Casini et al. 2017; Gary et al. 2013).  This is even more relevant now because of direct observations of the upper layers of the corona from space are now becoming available.  Currently the Parker Solar Probe (PSP) is obtaining *in situ* data on the coronal plasma and magnetic field at about 9 to 35 solar radii during its perihelion passages (c.f. Fox et al. 2016).  Faraday rotation measurements of the corona using the MESSENGER spacecraft carrier tone as it passed behind the sun (Kobelski et al. 2016; Jensen et al. 2018) fortuitously observed a CME that was launched across the line-of-sight.  From that the electron density, velocity, and magnetic field properties of the crossing plasma were well determined at ~ 6 solar radii above the surface.  All near surface data on the CMEs which intersect PSP orbit or MESSENGER line-of sight are important to the measurements of the corona.  Therefore, it is timely to introduce the capabilities of multi-wavelength, cm-λ (at 4 bands between 3.1 and 14.0

GHz) radio measurements available on the Goldstone Apple Valley Radio Telescope (GAVRT) with a demonstration of the 2012 eclipse observations of three solar active regions.

The GAVRT program operates a 34-m radio telescope with a wide-band, dual polarization low noise receiver. The Jet Propulsion Laboratory (JPL) and the Lewis Center for Educational Research (LCER) jointly conduct the GAVRT program as a way to introduce K-12 students to doing real science (see Roller & Klein. 2003). The GAVRT facility currently being used for making occasional maps of the Sun simultaneously at four cm-λ bands at 9.6, 5, 3.5 and 2.1 cm (3.1, 6, 8.4, 14 GHz) as part of the educational program. Solar Patrol is one of the study areas for the GAVRT program in a school curriculum. The annular eclipses of the Sun on 2012 May 20 observations were conducted remotely from the Senior Science Lab of Ribet Academy in Los Angeles through the GAVRT Michael J. Klein Control Center at LCER. This provided unique opportunities to do research, involve new students and teachers, assess DSS-28's suitability as a solar radio telescope, and make the GAVRT program better known.

**Why Eclipse Observations:**

GAVRT wavelength coverage at 9.6, 5, 3.5 and 2.1 cm, is ideally suited to probe coronal temperatures and magnetic fields using the gyroresonance mechanism (Kundu, 1965; also see Fig 6b in section 4). GAVRT has the capability to produce daily maps of the Sun at multiple times every 20-30 minutes. Although the angular resolution of the 34-m telescope at centimeter wavelengths (beam sizes 2.6 to 10 arcmin) is sufficient to just resolve the active regions (Fig. 1), these are not high enough to resolve adequately the structures within the active regions at all wavelengths. The eclipsing moon's limb provides high angular resolution (arcsec) independent of wavelength which is not feasible with single dish antennas. Since single dish telescopes have relatively low angular resolution (large antenna beam sizes which scale with wavelength, as shown in Table 1), it is not feasible to make observations with high angular resolution simultaneously at many wavelengths. However, in the special circumstances afforded by a solar eclipse, (see Fig. 2) the crossing of the moon's limb across an active region provides structural information in the direction perpendicular to the limb, with high angular resolution (limited only by the speed of moon's motion and the data sampling rate), simultaneously at all available wavelengths.

Solar eclipses played a pioneering role in the early years of radio astronomy. The 1948 solar eclipse provided the first opportunity to determine the precise location of the regions of enhanced emission in the solar corona (c.f. Orshiston, 2004). Kundu & Velusamy (1968) reported the cm-observations at 4cm wavelength of the solar eclipse of July 20, 1963, interpreting the source structure inferred from the strip scan in terms of gyroresonance absorption. Gary & Hurford (1987) used May 30, 1984 eclipse observations to spatially resolve microwave observations of a complex active region at 16 frequencies in the range 1.4-8 GHz. The strip scans across active regions provided by the crossing of the moon's limb helped delineate the change in the emission characteristics from free-free (also called bremsstrahlung) in the dense loops to circularly polarized gyrorésonance emission in the regions of intense magnetic fields and high brightness temperatures typical of greater heights in the corona than that for free-free emission at lower heights (see Fig. 6b).

There is a growing community interest in solar eclipse observations. For example, the Society of Amateur Radio Astronomers (SARA: *http://www.radio-astronomy.org/node/142*) offers community support, recognizing that a solar eclipse provides us with a rare opportunity to engage students by exploring many different aspects of solar radiation. In particular observing at radio wavelengths is often more interesting to students who are fascinated by things that they can't *"see"* and provides an opportunity to cultivate their interest in science.

One of our objectives, in this paper, is to demonstrate how solar eclipse observations can provide an opportunity to motivate and sustain their science interests at the same time generating useful data for the scientific community. Our main objectives are (i) to give an example of how GAVRT program and K-12 student/teacher participation can produce science data useful to the scientific community and science missions; (ii) to demonstrate the capability of the GAVRT telescope (NASA's DSN 34-m antenna) for Solar Patrol, and (iii) to present the results of the GAVRT observations of the 2012 solar eclipse of active regions and interpret the cm-$\lambda$ data at all 4 bands in terms of the gyroresonance mechanism probing the magnetic field and temperatures in the corona. The details of the GAVRT antenna and the receiver system for solar patrol are presented in Kuiper et al. (2020). Here we focus on the analysis of the eclipse observations.

## 2. Observations

GAVRT program nominally uses two of the Deep Space Network (DSN) antennas designated as DSS-13 and DSS-28. Generally, DSS-13 is used for Jupiter monitoring Jupiter's flux density at 2.28 and 8.45 GHz while the DSS-28 is being used for Solar Patrol observations. DSS-28 was designed for operation from 0.5 to 14 GHz (Imbriali et al. 2011; Jones et al. 2010). However, it has since been determined that the Low Frequency Feed is not usable due to intense radio interference. Therefore, for solar patrol observations it is used only at high frequencies > 3 GHz. The antenna and receiver parameters used for the eclipse observations are listed in Table 1. The eclipse observations used simultaneously four frequency bands between 3 and 14 GHz. For each band there were two receiver outputs corresponding to the right and left circular polarizations (RCP & LCP). For the eclipse observations we used 110 MHz bandwidth. For each band there were two receiver outputs corresponding to the right and left circular polarizations (RCP & LCP). The two circularly polarized outputs from the front end are converted to an intermediate frequency (IF) centered at 22 GHz. This is then mixed in quadrature (two IFs separated in phase by 90º to baseband (centered at 0 Hz). These are combined to form upper and lower sidebands. The upper sideband is then filtered to provide 110 MHz of bandwidth centered at 320 MHz. The AC signal from the filter is rectified with a tunnel diode. The DC voltage is then converted to a frequency, and the resulting cycles are counted for the specified integration time. The recorded data are then in units of "counts". Thus, the counts are a measure of the solar flux collected by the antenna, in each polarization.

In preparation for the eclipse a few sets of raster scan maps were observed simultaneously at all 4 wavelength bands (see Kuiper et al. 2020) and one set of maps observed just 2 hr before the eclipse is shown in Figure 1. These maps were observed simultaneously at all 4 bands with identical receiver settings that were used to observe the solar eclipse two hours later. Since the receivers were not designed for signals as strong as that of the sun, the attenuators were mostly at their maximum setting during the mapping and eclipse observations. Unfortunately, the attenuators were not well calibrated. Therefore, we did not follow the nominal calibration procedure (see Kuiper et al. 2020). Instead, we adopted a more realistic approach for calibrating the channel gain factors (conversion from data in counts to brightness temperature, $T_b$) using the quiet Sun, free-free emission, brightness temperatures as listed in Table 1, estimated from data compiled by Fürst (1980).

**Table 1 GAVRT Observational parameters.**

| Band GHz | Wavelength cm | Polarization LCP/RCP | FWHM arcmin | Quiet Sun** $T_B$ (K) | Map rms 1-σ (K) | Scan rms 1-σ (K) |
|---|---|---|---|---|---|---|
| 3.1 | 9.67 | R/L | 9.76 | 40000 | 90 | 52000 |
| 6 | 5.00 | R/L | 5.07 | 22000 | 27 | 8200 |
| 8.45 | 3.55 | R/L | 3.82 | 18000 | 12 | 2800 |
| 14 | 2.14 | R/L | 2.68 | 10000 | 15 | 2400 |

** due to thermal bremsstrahlung (free-free) emission

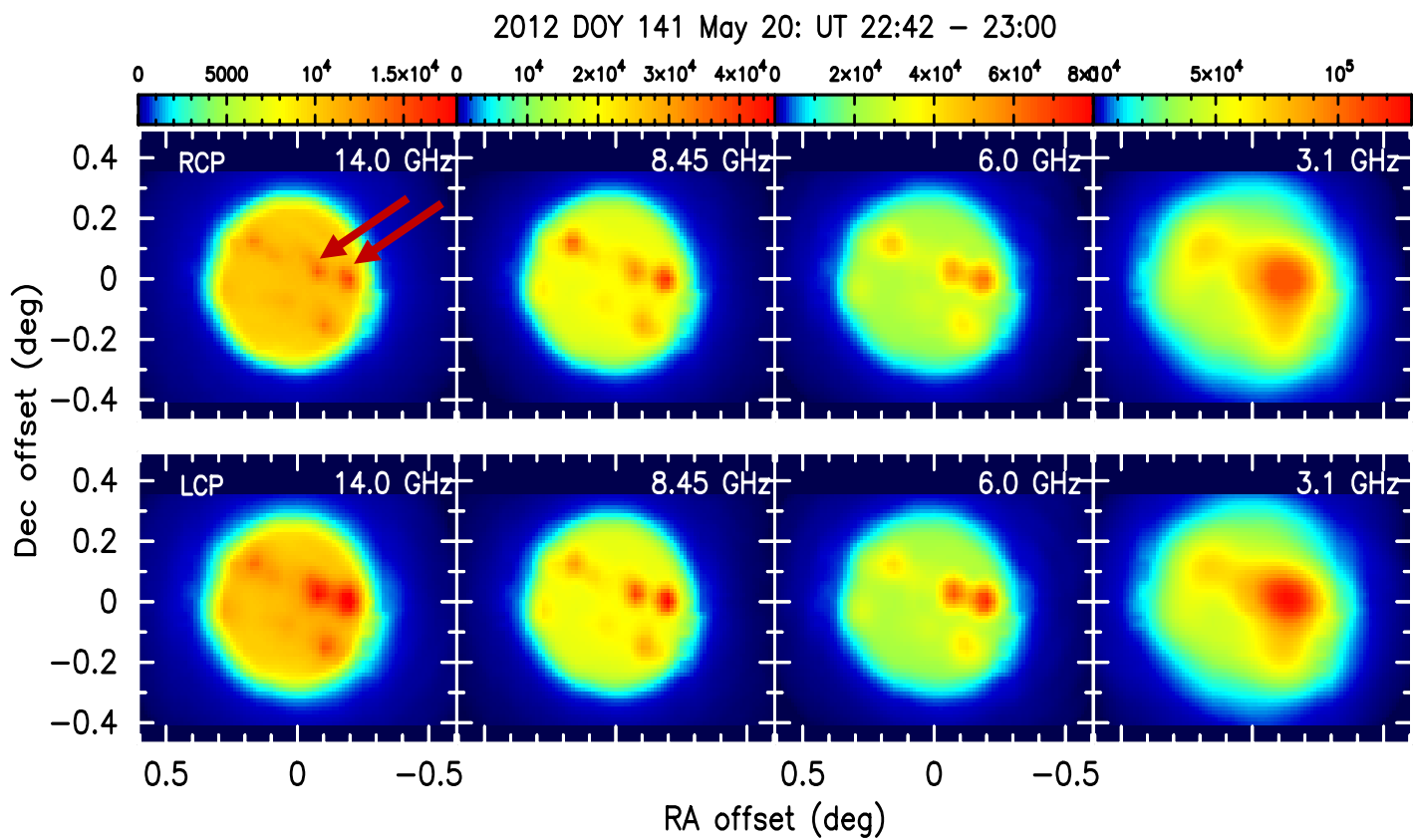

**Fig. 1** GAVRT maps of the Sun observed on May 20, 2012 two hr before the eclipse. The arrows on top left panel mark the two active regions used as targets for the eclipse. In all bands the emission surrounding the bright active regions (distinguished by deep red) is likely due to thermal bremsstrahlung from coronal structures at lower heights (see Fig. 6b).

In the absence of active regions at these frequencies sun's emission is due to thermal bremsstrahlung (free-free) which is unpolarized. Therefore, over the quiet sun region of the map both RCP and LCP maps have identically equal brightness temperatures. The raster scan intensity maps for each polarization and band were first made in units of the digitized output counts. In Fig. 1 in all bands the active regions are easily identified (distinguished by deep red). Thus, a quiet sun region that is devoid of any active region was easily identified. The channel gain factors (conversion from data in counts to brightness temperature, $T_b$) were obtained by matching the map intensity (in counts) to the known quiet sun brightness temperatures. Using these channel gains the intensities in the final maps were converted to brightness temperatures. In examining the maps (Fig. 1), it should be kept in mind that, away from active regions, the sun should have roughly equal brightness temperatures in LCP and RCP. However, it is quite usual for active regions to show strong circular polarization, which can be in either sense, depending on the geometry of the magnetic field and the wave propagation effects of the ordinary (o-) & extraordinary (x-) modes through the magnetized plasma above the sun spots (see section 4).

The scientific objective was to track an active region first when it was being eclipsed and later being exposed as it emerged from behind the moon and to use the moon's limb as a knife-edge to observe small-scale structure in the active regions (c.f. Kundu & Velusamy, 1968). For the 2012 annular eclipse we identified two complexes, AR1 containing the AR 11479 and 11482, and AR2 containing AR 11484 as marked in Figure 1. While antenna was tracking the active regions, the digitized output (counts) for each channel was sampled at 0.1s interval. The raw data (*eclipse curve*) for each band (RCP & LCP channels) are displayed in Figure 2 as counts plotted against time (UT) during the immersion (ingress) and emersion (egress) phases of the eclipse

The eclipse observation started around 00:20 UT about 15 minutes before the predicted time of eclipse of the active region AR1 when the antenna was tracking AR1 and the data (counts) were being recorded. In Fig. 2b we show the data from 00:27 UT when the counts were high as the region was fully in view. Figs. 2a & c show antenna beams at all GAVRT bands and the motion of the moon's limb across the regions during the immersion and emersion, respectively. Only the beams covering region AR1 are shown and, for clarity, the coverage on AR2, when the antenna was repointed, is not shown. The arrows indicate the scan directions as the regions are being covered or uncovered by the motion of the lunar disk. For example, the corresponding decrease in the count rate (immersion eclipse curve) as parts of the active region AR1 are blocked by the moon is shown in left half of the plot in Fig. 2b. When the region AR1 was entirely covered by the moon and the count rate reached a plateau, the antenna was moved to point on active region AR2 (time indicated by the vertical dashed line). The right half of the plot in Fig. 2b shows the immersion eclipse curve for region AR2. For about 30 minutes both AR1 and AR2 were behind lunar disk and as shown in Fig. 2c at 01:30:00 UT region AR1 began to emerge. At this time the antenna had been repointed and was tracking region AR1. The left half of the plot in Fig. 2d shows the emersion eclipse curve increasing in count rate as parts of the region AR1 were being uncovered by the motion of the moon's limb. Once the emersion of region AR1 was complete the antenna was repointed and tracking AR2. The emersion eclipse curve is shown in Fig. 2d (right half). Thus, we were able to observe both the immersion and emersion of two regions. Unfortunately, the antenna elevation was below 8.5° for the last few minutes for the eclipse observations (see section 3).

Note that the receiver set up was identical during both the map and the eclipse observations for all bands. Therefore, to convert the counts, shown in Fig. 2 to brightness temperatures we used the same channel gain calibration used for the producing maps shown in Fig. 1. The primary focus of this observation is exploiting the high angular resolution achieved by the eclipse to study the structure across the active region and therefore, calibrating the relative brightness of the active region with respect to the quiet sun brightness as done for mapping, provides reasonable estimates, adequate for the interpretations presented here.

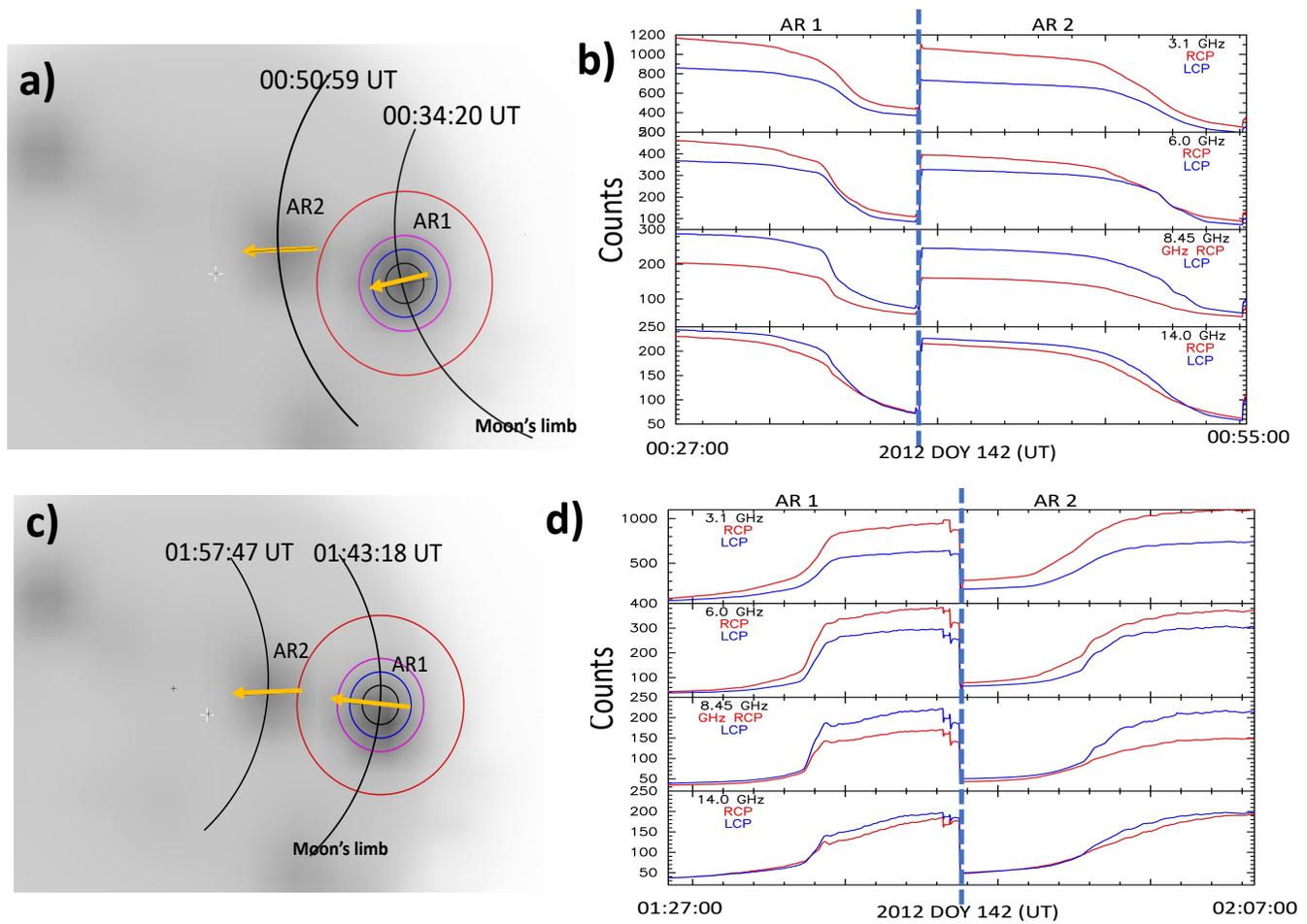

**Fig 2 (a) & (c)** Eclipse geometry during immersion (a) and emersion (c). Geometry of Moon's limb crossing the active regions (AR 1: 11479-82 complex; AR2: 11484) overlaid on the 8.45 GHz total intensity map in Fig. 1. The circles represent the FWHM beam sizes for 14 GHz (black), 8.45 GHz (blue), 6 GHz (magenta), 3.1 GHz (red). The arrow indicates scan direction of the eclipse. The arcs represent moon's limb at two selected times during the immersion and emersion. **(b) & (d)** Eclipse data (counts) for all bands during the immersion (b) and emersion (d) of active regions AR1 & AR2. The vertical dashed lines indicate the antenna pointing changed from region AR1 to AR2. The counts are a measure of the solar flux density in each polarization collected by the antenna.

## 3. Results

Figure 2 summarizes the geometry of the moon's limb crossing the active regions and the corresponding change in raw data counts. During the immersion (Fig 2a) the count rate decreases as parts of the active regions are blocked by the moon's disk. During the emersion (Fig. 2b) the count rate increases as parts of the active region emerge from behind the moon's disk. Working in the topocentric coordinate frame centered on the active region (beam center being tracked) we can then express the eclipse curve (see panels (b) in Fig. 2) in which the count rate is now a function of angular distance of moon's limb from the center of the active region.

The eclipse data (curve) was sampled at 0.1 sec interval, meaning each sample in the time series corresponds to 0.05 arcsec of sun being covered (immersion) or uncovered (emersion). The derivative of the eclipse curve yields the eclipse scan data which is a measure of flux density swept up by the limb within an arcsec wide (along scan direction) fan beam which is FWHM beam size long perpendicular to scan direction. In this way, the derivatives in units of counts become a proxy for brightness temperature. Since the raw derivatives are very noisy, they have been smoothed to an equivalent of 1~arcsec of the moon's motion across the sun. The brightness temperature scans derived from the eclipse curves are shown in the lower panels in Fig. 3. Counts-to-brightness temperature conversion used the gain factors derived from the quiet sun brightness for calibration as discussed above in section 2. However, these gain factors correspond to brightness averaged over the area of beams as marked in Fig 2, that is, the brightness of source size smaller than the beam size are reduced from the true value due to beam dilution (see below). Therefore, for the eclipse scan data we applied a correction for beam dilution considering the narrow fan beam for the eclipse strip scan data. The brightness temperatures in Fig. 3 and in all subsequent presentations use the brightness as would be observed by the 1-arcsec wide fan beam for each band. The 1-$\sigma$ uncertainty in map data and the corresponding uncertainty in the eclipse scan data are listed in Table 1.

The eclipse strip scan brightness temperatures are still lower than the true brightness temperatures by factors of few tens because the effective beam size is large perpendicular to the scan direction. For example, at 3.1 GHz the effective beam size for eclipse scan data is 1 arcsec x 9.8 arcmin. The 3.1 GHz source sizes are ~30 arcsec (see Figure 5) which is smaller than the beam size, perpendicular to the scan direction, by a factor ~20. Thus, the true brightness at 3.1 GHz is likely to be ~ 20 times brighter than those shown in the Figures. Because the source sizes (Figs. 3 & 4) and beam sizes (Table 1) scale with wavelength, we expect the true brightness temperature at all four frequencies likely to much higher by factors ~20 than those shown in the figures. Such values are consistent with the temperatures observed with high angular resolutions (Kundu & Velusamy, 1980; Kundu et al. 1981). The large-scale brightness temperatures at regions outside the active region (quiet sun) are not subject to the beam dilution.

In Figures 3 & 4 we present the observed strip scan brightness temperatures derived from the eclipse data for all three active regions, along with the Big Bear Solar Observatory (BBSO) H$\alpha$ and magnetogram images taken on May 21, 2012 (DOY 141) at 00:31:19 UT and 00:22:03 UT, respectively, the closest to time of the solar eclipse. The H$\alpha$ image was taken soon after the eclipse started as the moon's limb is seen over the sun's disk. We present RCP and LCP scan data separately because it is probable that the right and left circularly polarized emissions observed from earth may have originated at different layers above the sun spot due to differences in their optical depths (see section 4). The cm-$\lambda$ emission in the quiet sun region surrounding an active region is produced by thermal bremsstrahlung with corresponding brightness temperatures of $10^4$ to $10^5$ K. The gyroresonance absorption process is effective in the region above the sunspots (active region), and thus increases the brightness temperatures to above $10^6$ K and is observable as relatively compact bright source against a cool thermal background of the quiet sun. The corresponding sources have angular sizes ≤ 30 arcsec and have been observed at cm-$\lambda$ with high-resolution imaging instruments such as the Very Large Array (VLA) and Extended Owens Valley Solar Array (EOVSA) (e.g., McConell & Kundu, 1984; Lang, Wilson & Gaizauskas, 1993; Lee et al. 1999).

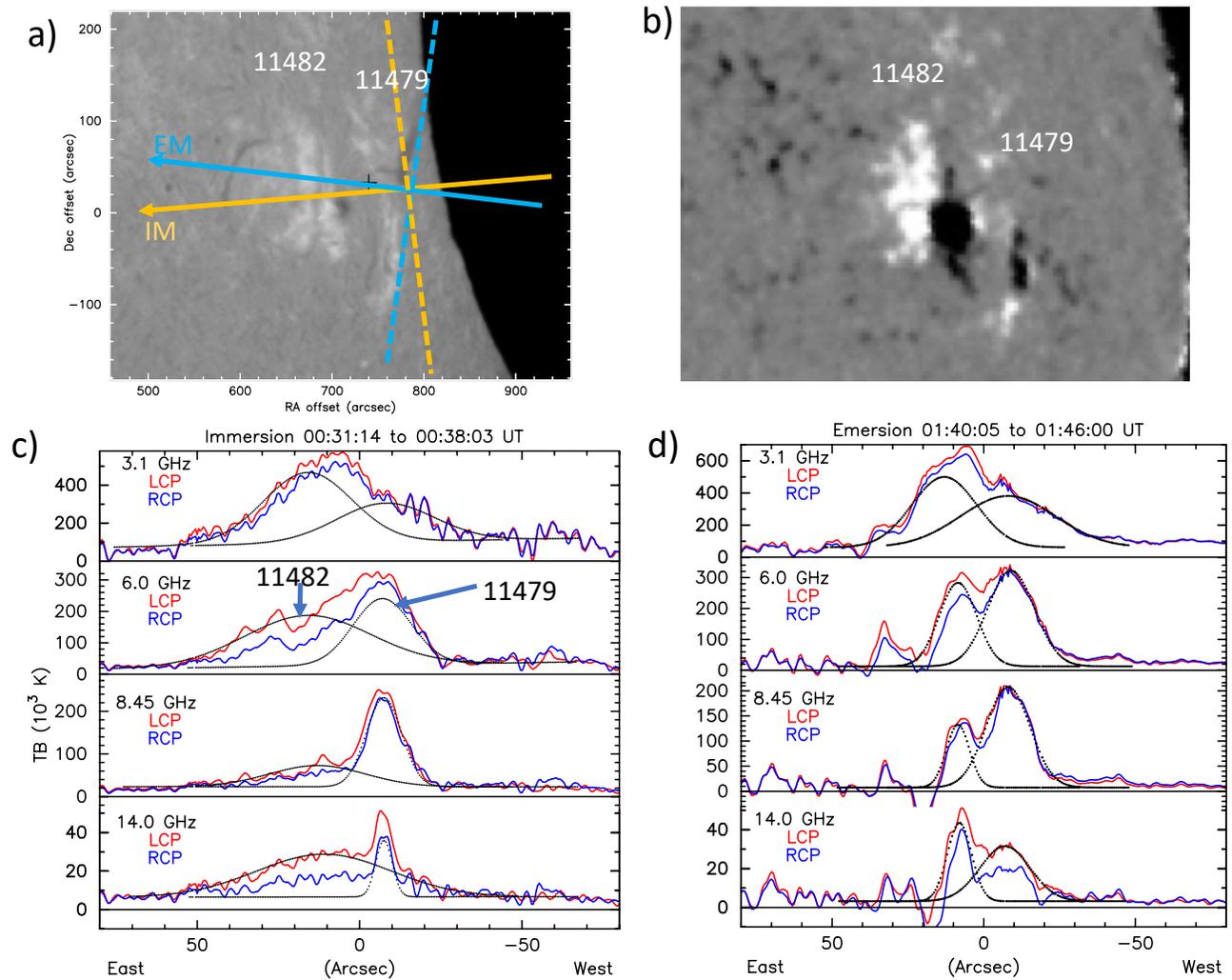

**Fig. 3** Brightness temperature, scans across AR1 region. The geometry of the "fan beam" orientation and the scanning directions are overlaid on BBSO Hα and magnetogram images of the active region AR 11479-82 taken on May 21, 2012 (DOY 141). **(a)** Hα image at 00:31:19 UT; **(b)** magnetogram image at 00:22:03 UT. The arrows mark the scan direction and the dashed lines indicate 1 arcsec wide "fan beam" (see moon's limb in Fig. 2). Lower panels show the RCP (blue) & LCP (red) brightness temperature scans across the active region AR1: 11479-82 at all 4 bands derived from the eclipse data presented in Fig. 2: **(c)** immersion, and **(d)** emersion. Black dotted lines indicate profile fitting to the LCP data with the two spatially resolved Gaussians corresponding to regions 11479 & 11482. For clarity the fitted profiles for RCP data are not shown (see Figure 5 for the results).

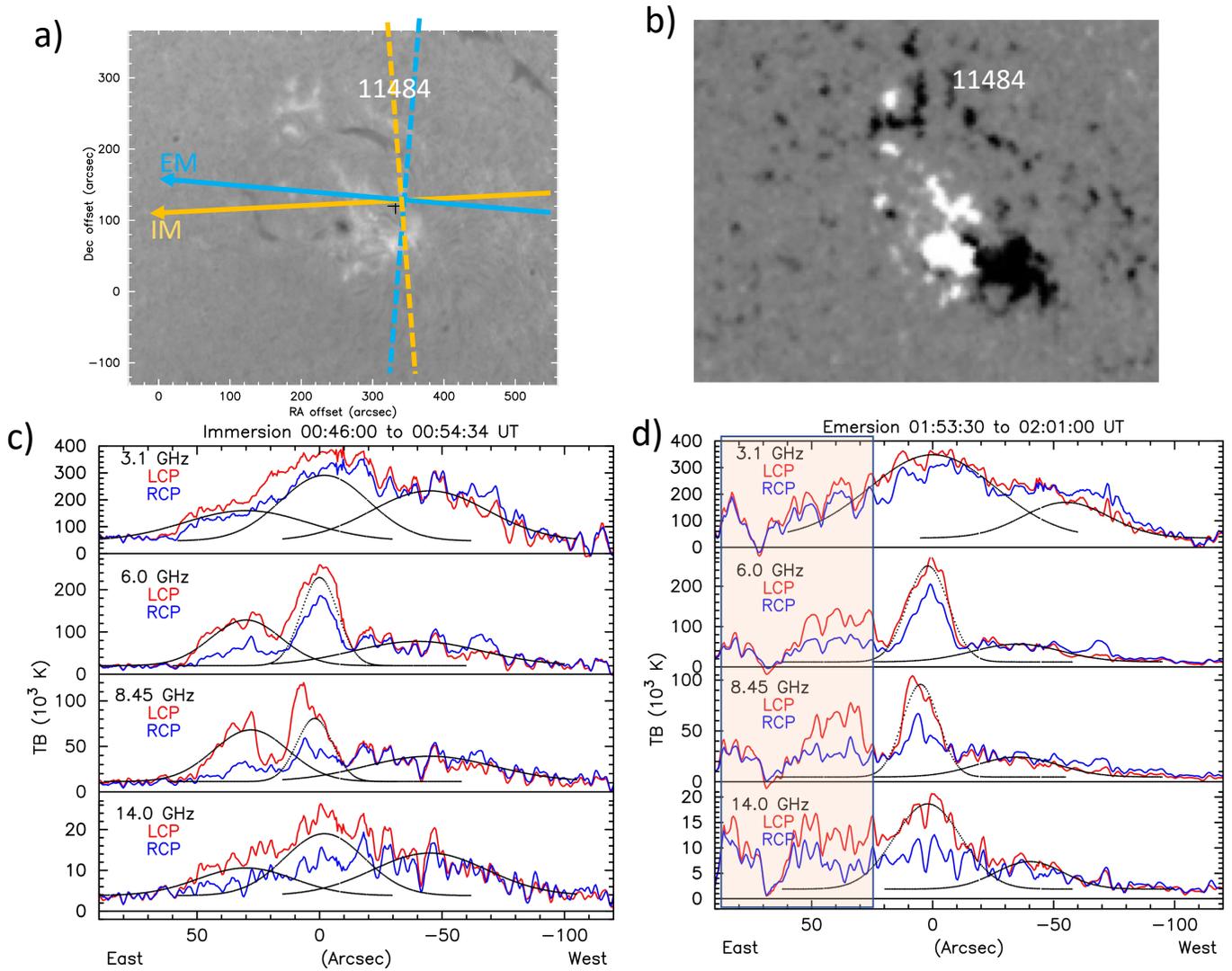

**Fig. 4** Brightness temperature scans across AR2 (11484) region. Caption same as in Fig. 3. The shaded area marks the observation taken when the antenna elevation was below 8.5°

The source size of the gyroresonance emission layer as a function of frequency depends on the configuration the coronal magnetic field above the sunspot. As seen in the data from BBSO magnetograms in Figs. 3b & 4b, all the active regions visible at the time of the eclipse had a bipolar structure at the sun's surface. Figure 3b shows two bipolar active regions. The separation between the centers of the two polarities is about half an arcminute for AR11482 and an arcminute for AR11484, which is a fraction of the radio telescope beam even at highest frequency 14 GHz (see Table~1). Therefore, the each of two polarizations is an average of the propagation modes above the two polarities. Without solving for the magnetic field in the corona from the magnetogram we assume that the two polarities are connected by closed field lines close to the surface and that at the heights of the radio emission the field has one polarity as revealed by the polarity of the observed radio emission.

As discussed in section 4, the field geometry ~~of a rapidly~~ with broadening of the isogauss surface in the corona would differ remarkably with height implying smaller source diameter at the base (observable at high frequency, 14 GHz) than high up in the corona (observable at low frequency, 3.1 GHz). Akhmedov et al (1983) carried out multifrequency (7.5 to 15 GHz) measurements using the RATAN-600 instrument. The high angular resolution in our present eclipse scan data also provides a wider frequency range to study the source structure and sizes. The temperature profiles in Figs. 3 & 4 show systemic variation from narrow to broad features at the highest (14 GHz) to the lowest (3.1 GHz) frequencies. The shapes of the derived scan profiles (see Figures 3 and 4) are complex which usually indicates the presence of several simpler shapes. In the absence of a theory to suggest the simple functional form we adopt the common practice of assuming a super position of Gaussians. For a quantitative analysis we fit multiple Gaussian components to the scan data estimating the source peak brightness and FWHM size.

The Gaussian profiles fitted to LCP data are overlaid as dotted (black) lines. For clarity we do not show the Gaussian fits to the RCP data, but the fitted peak brightness temperatures and the FWHM sizes are summarized in Fig. 5 for both RCP & LCP scans. To put our eclipse results in this perspective, in Fig. 5 we present a summary of the Gaussfit data, for all active regions, for both polarizations, and including both the immersion and emersion times. The results are shown as plots of brightness temperatures and source sizes as function of GAVRT frequencies in separate panels for AR 11479, AR 11482 and AR 11484 main component. Because of the complexity in the scan data for AR 11484 only the parameters for the main feature (see Fig. 4) are used in Fig. 5. The immersion and emersion times are distinguished by symbols cross and stars respectively. Likewise, the colors indicate RCP (blue) and LCP (red). Both the peak brightness and source size show strong dependence on GAVRT band frequencies and are discussed in detail in section 4.

It is somewhat surprising to note significant differences between the scan profiles in Figs. 3 & 4 between the immersion and emersion times. However, it is not improbable because: (i) though the scan directions differ only ~ 15º the differences in the scan profiles imply very complex structure within the fan beam which is extended perpendicular to scan direction; and/or (ii) temporal changes occurring between the immersion and emersion times separated by ~ 1.5 hr. Active regions are known to vary quickly on time scales of minutes, and slowly on time scales from hours to several days (c.f. Bogod & Tokhchukova, 2003).

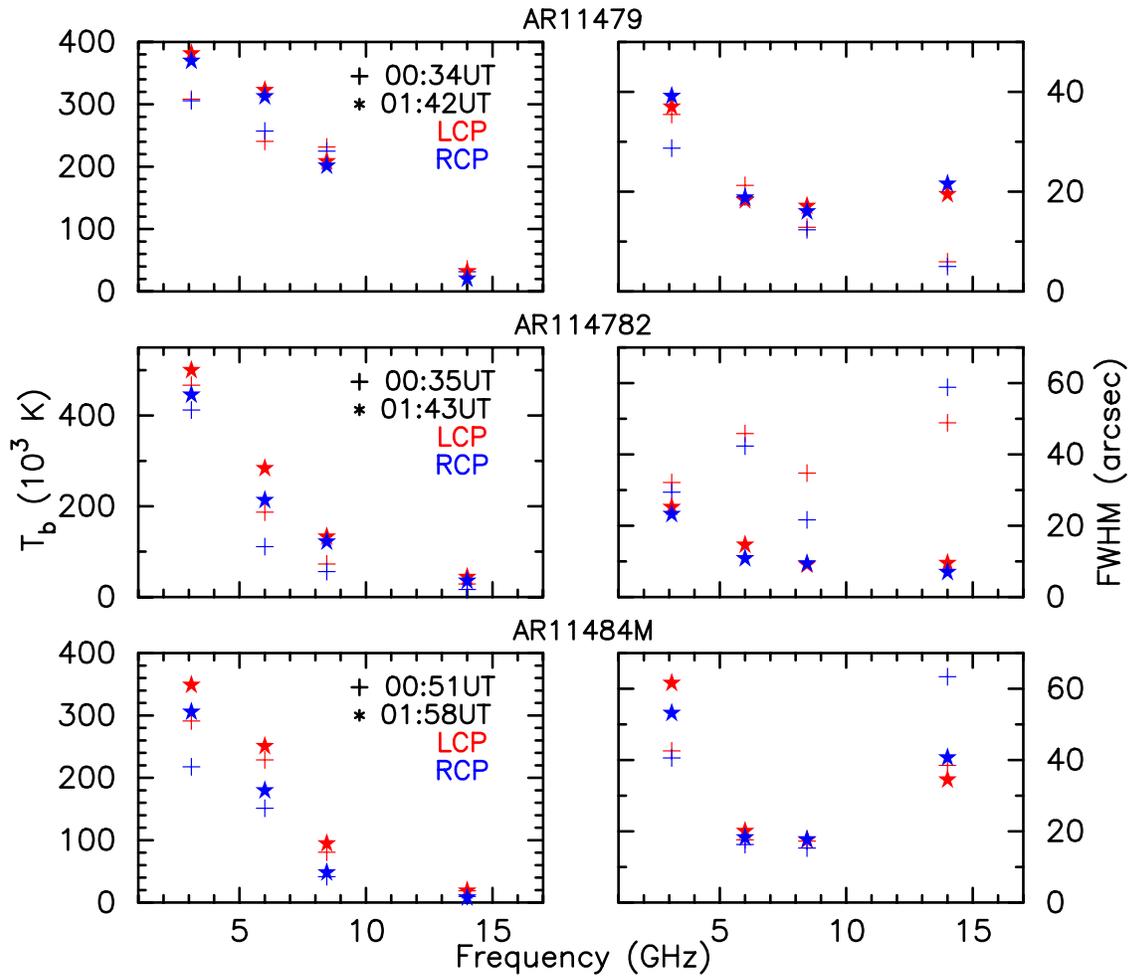

**Fig. 5** cm-λ brightness temperatures *(left)* and source sizes *(right)* as function of GAVRT frequencies for AR 11479 *(top)*; AR 11482 *(middle)* and AR11484 main component *(lower)*. The symbols cross and stars represent data for immersion and emersion times respectively. The colors indicate RCP (blue) and LCP (red). Large increase in angular size with height (wavelength) implies rapidly diverging magnetic field. Note that the true brightness temperatures are much higher when beam dilution is considered (see text).

## 4. Discussion

As discussed above, the strip scans across the active regions with arcsec resolution show a systematic increase in the source sizes (emission region) with increasing wavelength, in other words with increasing height into the corona. (See the Gauss fits to the emission profiles across the active regions in Figs 3 & 4 and plots of gauss fit source brightness and size in Fig. 5). We interpret the GAVRT eclipse results and the observed morphology of the cm-λ emission above the active regions in terms of the geometry of the magnetic field emerging from the surface into the corona and the electrons in the plasma streaming along the field lines, emitting by gyroresonance process. To put this in the context, in Fig. 6a we illustrate the temperature and density structure as a function of height above sun's surface. In the region at about 2000 km the properties of the solar atmosphere change drastically. This also divides the lower chromosphere from the upper chromosphere. The upper chromosphere blends smoothly into the corona where the temperature rises to a million K ($10^6$ K) and the electron density drops to about $10^9$ cm$^{-3}$. Fig. 6b identifies the region (height above the surface) which can be probed with radio waves in the range of 1 to 100 GHz for cases when the opacity (optical depth) is dominated by free-free emission, or by the gyroresonance absorption. For frequencies > 1 GHz the plasma layer is not important as it is well below the free-free absorption layer. However, for the cm-λ of the GAVRT bands, contributions to the optical depth from both free-free and gyroresonance are equally important. Nominally, in absence of strong magnetic fields (region above the quiet sun surface), the observed brightness temperatures correspond to that of free-free $\tau_{ff}$ = 1 layer. Furthermore, the effects of wave propagation in the magneto-ionic medium (Ratcliffe, 1962) are also important for characterizing the optical depths and polarization of the emerging radiation from any of these layers, as observed from earth. There are two independent modes of propagation in an magnetoionic medium, namely the ordinary (o) mode in which the effect of the magnetic field is less or absent, and the extraordinary (x) mode in which the effect of the magnetic field is greater, and they have different refractive indices and absorption coefficients.

~~However,~~ When high magnetic field is present (in the active regions above sunspots), as shown in Fig. 6b, the gyroresonance layers for s > 3 are clearly well above the free-free $\tau_{ff}$ = 1 layer. Thus, in contrast to thermal free-free emission, the gyoresonance absorption mechanism provides high enough optical depths to cm-λ emission higher up in the corona resulting in observing brightness temperatures > $10^6$ K. The gyrating electrons emit radiation at the gyro frequency and at its harmonic frequencies. For a given harmonic number *s* the gyro frequency $f_B$ is determined by the local magnetic field strength, B as

$$f_B (MHz) = 2.8 \times B \times s \quad (1)$$

where B is in gauss (e.g. White & Kundu, 1997). Typically, the cm-λ frequencies require an average strength of the magnetic field over the sunspot group in the range of about 600 gauss at a height of $2 \times 10^4$ km above the photosphere to 250 gauss at a height of $4 \times 10^4$ km. Thus, the relationship [Eq 1] between the frequencies of the GAVRT bands and the coronal magnetic field strength provides a set of unique localized emission layers Fig 6c shows schematic illustrations of magnetic field lines above an isolated sunspot and the gyroresonance emission layers for two GAVRT bands, 3.1 GHz & 8.45 GHz. The geometry of the magnetic field and plasma in an emerging magnetic flux region in a slowly evolving quasi-steady phase has been well studied (e.g. Forbes & Priest, 1984. Though in general the magnetograms show bipolar fields we can assume either i) that the two polarities are connected by closed field lines close to the surface and at the heights of the radio emission the field has one polarity or ii) because of the larger antenna beam size only the dominant polarity is observed. However, in the magnetogram AR11479 appears to have dominant single polarity and geometry in Fig. 6c seem to be consistent with the angular sizes of the isogauss surfaces as interpreted from gyroresonance emission as shown in Fig. 7.

As shown Fig. 6 the low harmonics requiring higher magnetic field strengths are located closer to the surface deeper into the corona while the higher harmonics (s=4) requiring B ~ 276 gauss for 3.1 GHz band is located at a greater height. Thus, the direct detection of cm-λ gyroresonance emission at any one of the GAVRT bands, is evidence for the presence of a corresponding magnetic field strength in the corona along the line of sight, provided the appropriate harmonic number is known (e.g., Hurford & Gary 1986; Holman 1992; Gary & Hurford 1994).

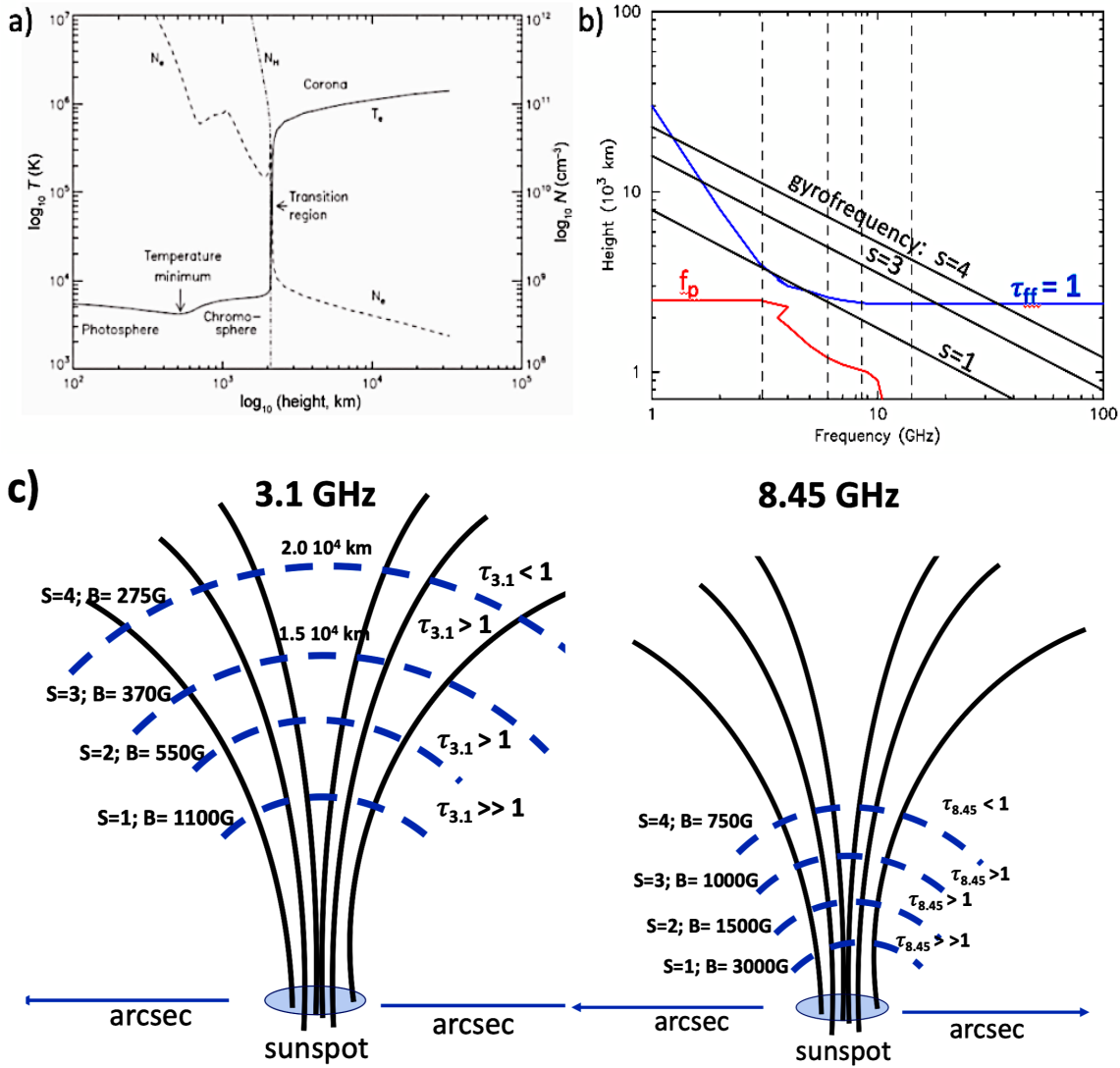

**Fig. 6 (a)** Temperature and density structure as a function height above the solar surface (from NASA *Cosmos*). **(b)** The gyroresonance frequency layers are identified in height-frequency space. For comparison the layers of plasma frequency $f_p$ (red) and free-free emission optical depth $\tau_{ff}= 1$ (blue) are also shown. (adapted from Gary & Hurford, 2004).

**c)** Schematic illustration of magnetic field lines above an isolated sunspot and the gyroresonsnce emission layers for two GAVRT bands, 3.1 GHz & 8.45 GHz., respectively. The dashed lines show the isogauss contours corresponding to the harmonics, s = 1 to 4, of the gyrofrequency appropriate for the selected GAVRT frequency. The dashed lines also identify the gyroresonance absorption layers shown schematically (not to scale) as function of height in the corona above the surface (adapted from Kundu & Vlahos, 1979 for the GAVRT cm-λ bands). Note this oversimplified magnetic field geometry, in general, may not be applicable to bipolar regions, except AR 11479 which appear to have dominant single polarity.

At cm-λ frequencies only the emission from the highest optically thick harmonic layer in the corona is observed, because all lower-lying layers (corresponding to higher B and hence smaller $s$) are obscured by the process of resonant absorption in the optically thick upper layers. Which of the harmonic frequencies corresponds to observed brightness at a given band is determined by optical depths to each harmonic frequency layer which have been well studied by several groups (e.g. Kakinuma, & Swarup, 1962; Zheleznyakov 1962; Zlotnik 1968a, 1968b). We can generally expect the observed circular polarization to coincide with the photospheric magnetic polarity, the RCP/ LCP emission appearing above the positive/negative magnetic polarity regions. However, for complex and extended active regions, the observed sense of circular polarization may sometimes disagree with the photospheric magnetic polarity distribution. The interpretation of such observations requires mode coupling theory.

Based on the observed brightness temperatures, we can associate the left and right circularly polarized waves with the x- and o-modes, respectively (see Lee et al. 1998b for details). Hereafter, we denote the brightness temperatures in the LCP and RCP polarization states as x- and o- modes, respectively while attributing them to specific harmonics of the gyro frequency. Generally, the x-mode is always more likely to be optically thick than the o-mode (c.f Zheleznyakov 1962). Recently, Lee et al. (1998a) calculated gyroresonance emission from magnetic field-line structured temperature models and using a magnetic field model calculated via a nonlinear force-free field extrapolation of photospheric vector magnetograms and compared them with high resolution VLA maps. They show that in general the brightest radio emission regions are optically thick up to the fourth harmonic for all frequencies (5 & 8 GHz) and both polarizations (x- & o- modes), with few exceptions of the o-mode being optically thin for s=4 and thick only in s=3.

Below, we show that the cm-λ source brightness temperatures and the morphology (angular size), derived from the eclipse data observed simultaneously at all for GAVRT bands, are consistent with the gyroresonance emission models. The schematic of the gyroresonance emission model shown in Fig. 6c can be used to interpret the observed cm-λ source brightness temperatures and sizes (isogauss surface) as function of GAVRT frequencies. For qualitative comparison with the eclipse scan results for active region AR 11479, 82 & 84 shown in Figures 3 & 4 we use this as a toy model. For demonstration purposes in Fig. 6c we show that s=4 layer approximately delineates the transition from optically thick to thin regimes for each band. In reality it can be the s=3 layer or even less. However, the emissions from layers for harmonics s > 5 are always negligible with the corresponding optical depth, $\tau \ll 1$. The over-simplified representation in Fig. 6c does not distinguish between the x- & o- modes. In real scenario the optical depths for a given harmonic and hence the absorption/emission layer can be very different for LCP (x-mode) and RCP (o-mode).

Our eclipse results of the cm-λ source brightness and size as function of GAVRT frequencies for AR 11479 at time of immersion are shown in Fig. 7. The fitted temperature profiles for both polarizations, LCP & RCP are shown in Fig. 7a , with the 3.1 GHz at top and 14 GHz at the bottom. Note the increasing source size with wavelength giving a perception of height, as inferred from the brightness temperature, illustrates the broadening of the isogauss surface of the magnetic field geometry. Note source sizes of the gyroresonance emission layers correspond to the isogauss surface layers shown in Fig. 6c. The Gaussian profiles and angular size (half-width) were taken from the fits to eclipse scan data in Fig. 3c.

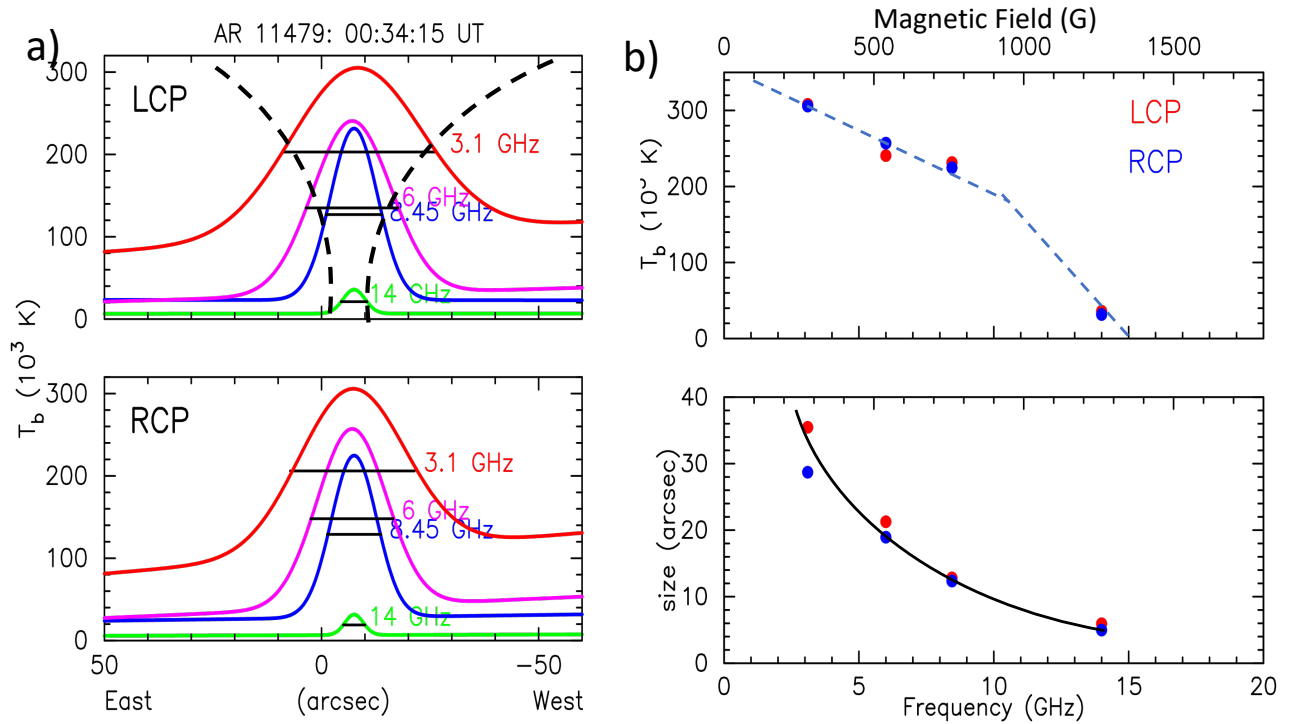

**Fig. 7 (a)** cm-λ Gauss fit profiles of brightness scans across AR 11479 at time of immersion are stacked as function of GAVRT frequencies for illustrating the sizes of isogauss surfaces of the magnetic field geometry, as measured by gyroresonance emission layers. The Gaussian data taken from the LCP & RCP eclipse scans in Fig. 3c. The bands are identified by the colors as marked. **(b)** Gauss fit peak brightness temperatures (top panel) and the source size (lower panel) are plotted as function of frequency. The magnetic field strengths required for gyroresonance 4$^{th}$ harmonic frequencies are indicated along the X-axis at top.

In Fig. 7b the fitted peak brightness temperatures (top panel) and the source size (lower panel) are plotted as function of frequency. The magnetic field strengths required for gyroresonance 4$^{th}$ harmonic frequencies that match the GAVRT bands are indicated along the X-axis at top. Plotting the eclipse scan results in a temperature — frequency space provides more insight into the temperature structure in the corona: (i) the brightness temperature variation with wavelength, as indicated by the dashed line resembles temperature —height distribution shown in Fig. 6a. (ii) The temperature-frequency variation can be translated to represent the relationship between the corona temperature and magnetic field strength above the active regions. Such data are very useful for understanding the role of magnetic field in heating the corona. (iii) In the case of AR 11479 both the RCP and LCP brightness temperatures are in good agreement suggesting both x- and o- modes for harmonic s=4 are optically thick, similar to the examples in the VLA maps (Lee et al. 1998a).

Similar to the temperature-frequency plot the eclipse data plotted in the size-frequency space is also equally informative about the magnetic field structure above the active region. The lower panel in Fig. 7b shows a strong wavelength dependence for size variation indicating increase in size with height above the active region which is consistent with widening of the isogauss surface of the field geometry as discussed above. Since angular size at a frequency band corresponds to the extent of the gyroresonance emission layer, it is also a measure of the extent of the isogauss surface with field strength corresponding the band frequency as marked at the top. As before we can now translate the functional form of the size—frequency relation (as indicated by the solid line) into isogauss surface width — magnetic field relation. Such size—frequency (size—magnetic field) relationship when combined with *a priori* data on temperature—height (Fig. 6a), temperature—frequency or temperature—magnetic field (Fig. 7b top panel) provides valuable constraint on the models/algorithms used to extrapolate surface magnetic fields to coronal heights (e.g. Casini et al. 2017).

In the temperature-frequency plots in Fig. 5 with the only exception of 6 GHz data for AR 11479 at the time of immersion, in all other cases the LCP brightness temperature greater than RCP temperatures, as may be expected because the optical depth $\tau_x > \tau_o$. In the case of AR 11482 and 11484 the LCP brightness temperatures are significantly higher than those of RCP. If we assume s = 4 for LCP emission layer, this difference suggests the s= 4 o-mode (RCP) is optically thin and RCP emission must include contributions from to s < 4 layers, deeper in the corona. It may be noted that here we assume that at these cm-λ wavelengths active regions exhibit monopolar characteristics because the surface field polarity does not extend into corona, as the field directly above the surface is more likely to be closed, thus likely to have only one direction in the corona. To further examine the possibilities of such situation, below we present an analysis of the results for AR 11484M at the time of immersion.

As an example, in Fig. 8 we show the data for AR 11484M- immersion. The cm-λ brightness temperatures as function of GAVRT frequencies and magnetic field inferred from the gyrofrequencies illustrate the possibilities that the LCP and RCP data are probing, respectively, the higher and lower levels in the corona. We show that the observed LCP/RCP brightness temperatures can be understood if we assume that the highest harmonic for LCP (x-mode), to be optically thick, as s=4 and for RCP (o-mode) to be < 4. In Fig. 8 for each LCP/RCP data point we identify the most likely harmonic number & the required magnetic field strength. Note that the harmonics of the gyroresonance frequency as marked were identified to be consistent with both the magnetic field variation and brightness temperature (or inferred height). The solid (black) lines represents the temperature —frequency profile which can also be interpreted as temperature —magnetic field profile if we consider the magnetic field associated with each data point, as marked. Note that in this schematic for magnetic field and density (the long arrows on the right) only the values at the extremities are shown and the true functional forms of their variations with height may not be linear.

Figure 8 clearly demonstrates how the cm-λ GAVRT data provides useful constraints on interpreting gyroresonance source models. The GAVRT eclipse data show that cm-λ radio data may be used as a diagnostic of coronal magnetic structure and temperature probes. Because the emission at these frequencies are due to a resonant mechanism, the optically thick emission arises from a relatively thin (typically 100 km in depth) surface of constant magnetic field strength (appropriate to the specific harmonic of the gyrofrequency). This has several advantages over other optically thin probes such as EUV which are emissions measured as weighted averages of all of the material in a resolution element along the line of sight. Generally, EUV data has no information on magnetic fields. However, Li et al (2016) have shown new opportunities using magnetic field induced transitions of two $Fe^{9+}$ fine structure lines as diagnostics of coronal magnetic fields.

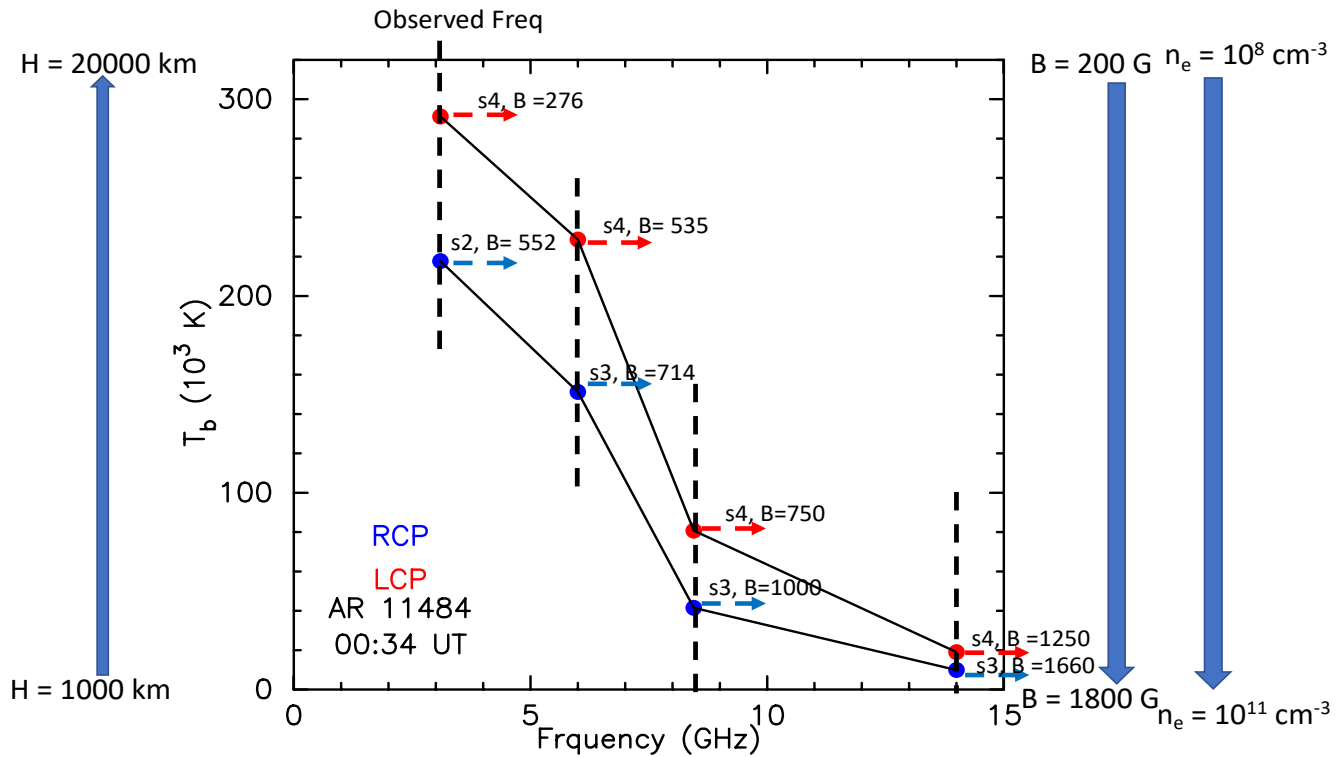

**Fig. 8** A schematic demonstrating which gyroresonance harmonic layers in the corona and corresponding magnetic fields are probed by the LCP and RCP data in the GAVRT frequency bands. The derived cm-λ brightness temperatures for AR 114784M (- immersion) are plotted as function of GAVRT frequencies. The colors indicate RCP (blue) and LCP (red) data. The short red and blue arrows identify the most likely harmonics & the required magnetic field strength for each LCP and RCP data points, respectively. The solid (black) lines represents the observed temperature profile for the gyroresonance harmonic number, s=4 (x-mode: LCP) and the s= 2 or 3 (o-mode RCP). The dashed vertical lines indicate the GAVRT frequencies. The long arrows to the left and right of the plot indicate, schematically, the ranges of the corona height above the surface, magnetic field strengths and electron densities that correspond to the range of the observed brightness temperatures (10 to 300 thousand K).

Large increase in angular size with wavelength (height) implies rapidly diverging magnetic field above mono-polar region or in general broadening of the isogauss surface in the corona above multipolar regions. The angular size at a given wavelength is a measure of the isogauss surface area appropriate to harmonic number of the gyrofrequency, and height above the surface appropriate to brightness temperature. With the exception of AR 11482 at the immersion time all other data show increase in source size from 8.45 to 3.1 GHz as expected for diverging field lines. However, the 14 GHz source size data is less conclusive partly due far-less contrast between the gyroresonance peak and the thermal background around it. As seen in Fig. 6b at frequencies above 10 GHz the free-free optical depth becomes comparable to gyroresonance absorption optical depths. The scan profile for AR 11482 at the immersion time seem more complex (see Fig. 3c) and it is possible that the individual gyroresonance peaks are not resolved and appear to be blended. Alternatively, the magnetic field lines are closed at base itself such that the isogauss contours have similar angular sizes, but located at different heights, consistent with the temperature— frequency/magnetic field plot.

As mentioned before, in the temperature—frequency plots, the true brightness temperatures are much higher when beam dilution is considered. Nevertheless, they indicate an overall trend on how the temperature and size change with frequency which will be useful in modeling the magnetic field geometry and the temperature & density structure above the active region. Furthermore, whenever it is possible to identify the harmonic number of the optically thick gyroresonance frequency we can easily translate the brightness temperature — frequency plots as brightness temperature—magnetic field plots which is important for the models/algorithms used to extrapolate surface magnetic fields to coronal heights (e.g. Casini et al. 2017). Thus, they provide magnetic field line connectivity as well as the field strength and its relationship to temperature structure in the corona (Schmahl et al. 1982; Alissandrakis & Kundu, 1984; Nindos et al. 1996; Lee e al. 1998a; Lee, 2007).

## 5. Conclusion

The results of our multiwavelength eclipse observations, presented here, demonstrate the value of high angular (arcsec) resolution achieved by the Moon's limb simultaneously at all wavelengths to derive the cm-$\lambda$ groresonance emission source brightness temperatures and source sizes. The wide frequency range (3.1 to 14 GHz) of the GAVRT eclipse data adequately sample and quantify the changes in brightness temperatures and source sizes as functions of frequency and enabling their interpretation as variation with height above the sun's surface. We present examples for diverging or broadening of the isogauss surface in the coronal magnetic field geometry above the active regions which are readily observed in the frequency—angular size (contours of isogauss surface) relationship. The data presented here also show evidence how the frequency – brightness temperature relationship is translated to a magnetic field – brightness temperature relationship. The eclipse data for one of the active regions at the time of immersion presented in Fig. 5 demonstrate the usefulness of similar data sets in a broader context of self-consistent modeling of the temperature and magnetic field structure. The unique property of gyroresonance cm-$\lambda$ emission will remain important for the improved coronal magnetic field models made feasible by recent progress in coronal field extrapolation, given the photospheric vector magnetogram measurements.

The GAVRT team is continuing its effort to include student/teacher participation in Solar Patrol program. In this paper we presented a frame work for analyzing the data observed under the GAVRT program designed for Solar Patrol with teacher-student participation. Our results demonstrate the usefulness of the data taken by GAVRT for frontier science research, especially in the context of the current interests driven by the Parker Solar Probe mission. Furthermore, one of our objectives is that the results presented here will serve as an example to sustain the growing educational community interest in solar eclipse observations.

**Acknowledgement:** We thank referee Dr, Philip Judge for helpful comments. We thank Dr Jay Pasachoff for suggesting GAVRT observations of the 2012 solar eclipse. We also thank Dr. Dale Gary for his help with planning the GAVRT observations and for proving the pointing and tracking information for the active regions. We acknowledge the support of Lisa Lamb, President/CEO and LCER staff for all their infrastructure & observational support to data presented in this paper. We thank Dr. Joseph Lazio for critical comments. This work was performed at the Jet Propulsion Laboratory, California Institute of Technology, under contract with the National Aeronautics and Space Administration.**REFERENCES:**

Akhmedov, S. B., Gelfreikh, G. B., Bogod, V.M. & Korzhavin, A. N., 1982, Sol. Phys. 79, 41

Alissandrakis, C. E., & Kundu, M. R. 1984, A&A, 139, 271

Bogod, V. M. & Tokhchukova, S. Kh., (2003) ", Astronomy Letters, 29, 263

Casini, R., White, S. M., Judge, P. G; (2017),", Space Sci Rev, 210, 145

Fox, N. J., Velli, M. C., Bale, S. D., et al., 2016, Space Sci Rev 204, 7

Fürst, E., 1980, in IAUS 86: *Radio Physics of the Sun*, p25

Gary, D. E., G.D. Fleishman, G. D, Nita, G. M., 2013, Sol Phys. 288, 549 (2013)

Gary, D. E. & Hurford, G. J., 2004, "Radio Spectral Diagnostics", in Solar Space weather Radiophysics, Vol 303, pp71-87.

Gary, D. E. & Hurford, G. J., 1987, ApJ, 317, 522

Gary, D. E., & Hurford, G. J. 1994, ApJ, 420, 903

Forbes, T. G. & Priest, E. R., 1984, Solar Phys. 94,315

Holman, G. D. 1992, in Proc. First SOHO Workshop: Coronal Streamers, Coronal Loops, and Coronal and Solar Wind Composition, ed. C. Mattok (Paris: ESA), 189

Hurford, G. J., & Gary, D. E. 1986, in Coronal and Prominence Plasmas, ed. A. I. Poland (NASA Conference Proc. 2442; Washington NASA), 319

Imbriali, W. A., Weinreb, S., Jones, G., Mani, H. & Akgiray, A., 2011, , IEEE Transactions on Antennas and Propagation, 59, 1954

Jensen, E. A.; Heiles, C. Wexler, D., Kepley, A. A., Kuiper, T. B. H., et al., 2018; *Ap.J.* 861, 118

Jones, G., Weinreb, S., Mani, H., Smith, S., Teitelbaum, L., Hofstadter, M., Kuiper, T. B. H., Imbriale, W. A., Dorcey, R & Leflang, J., 2010, Proceedings of the SPIE. Ground-based and Airborne Telescopes III, 7733, 773330

Kakinuma, T. & G. Swarup, G., 1962, Ap J. 136, 975

Kobelski, A.; Jensen, E.; Wexler, D.; Heiles, C.; Kepley, A.; Kuiper, T.; Bisi, M., 2016, ASP Conference Series, Vol. 504., p.99

Kuiper et al., 2020, in preparation

Kundu, M. R., 1965, *"Solar Radio Astronomy",* (New York: Wiley Interscience)

Kundu, M. R., (1982), "*Advances in Solar Radio Astronomy*", Rep. Prog. Phys. 45,1435

Kundu, M. R., Velusamy, T., 1968, Nature, 217, 1132

Kundu, M. R., Velusamy, T., 1980, ApJ. (Letters) 240, L63-L67

Kundu, M. R. & Vlahos, L., 1979, ApJ,232,595

Kundu, M. R., Schmahl, E. J. & Rao, A. P., 1981 Astron. Astrophys. 94, 72

Lang, K. R., Willson, R. F., & Gaizauskas, V., 1983, Ap J. 267, 455

Lee, J., 2007, Space Sci Rev 133, 73

Lee, J., McClymont, A. N., Mikiˊc, Z., White, S. M., & Kundu, M. R. 1998a, ApJ, 501, 853

Lee, J., White, S. M., Kundu, M. R. Mikiˊc, Z. & McClymont, A. N.,. 1998b, Solar Phys. 180,193

Lee, J., White, S. M., Kundu, M. R. Mikiˊc, Z. & McClymont, A. N., 1999, ApJ, 510, 413

Li, W., Yang, Y., et al. 2016, ApJ. 826, 219

McConell, D. & Kundu, M. R., 1984, ApJ, 279, 421
20